\documentclass[]{article}

\usepackage{amssymb}
\usepackage{cite} 
\usepackage{url}  
\usepackage{ifthen}  
\usepackage{multicol}   
\usepackage{geometry} 
\usepackage{amsmath}
\usepackage{array}
\usepackage{graphicx}
\usepackage[pdftex]{hyperref}
\usepackage[usenames]{color}
\usepackage{fancyhdr}
\usepackage{verbatim}
\usepackage{setspace}
\usepackage{graphicx}
\usepackage{tikz}
\usepackage{amsfonts}
\usepackage{bm}
\usepackage{mathtools}
\RequirePackage{hyperref}

\newcommand{\xb}{\mathbf{x}}
\newcommand{\yb}{\mathbf{y}}
\newcommand{\ub}{\mathbf{u}}
\newcommand{\Hb}{\mathbf{H}}

\begin{document}

\title{Algebraic identification of the effective connectivity of constrained geometric network models of neural signaling}
\author{Marius Buibas and Gabriel A. Silva\footnote{ Corresponding author: Prof. GA Silva\newline UC San Diego Jacobs Retina Center\newline 9415 Campus Point Drive\newline La Jolla, California 92037-0946\newline \newline Email: gsilva@ucsd.edu}\\Department of Bioengineeirng\\University of California, San Diego}
\date{}
\maketitle
\begin{abstract}
Cellular neural circuit and networks consisting of interconnected neurons and glia are ultimately responsible for the information processing associated with information processing in the brain. While there are major efforts aimed at mapping the structural and (electro)physiological connectivity of brain networks, such as the White House BRAIN Initiative aimed at the development of neurotechnologies capable of high density neural recordings, theoretical and computational methods for analyzing and making sense of all this data seem to be further behind. Here, we propose and provide a summary of an approach for calculating effective connectivity from experimental observations of neuronal network activity. The proposed method operates on network-level data, makes use of all relevant prior knowledge, such as dynamical models of individual cells in the network and the physical structural connectivity of the network, and is broadly applicable to large classes of biological and non-biological networks.
\end{abstract}

\newpage
\section{Introduction}
Cellular neural circuit and networks consisting of interconnected neurons and glia are ultimately responsible for the information processing associated with information processing in the brain.  While the entire network can range in size from a few hundred neurons, as in the case of the nematode worm \emph{C. Elegans}, to several hundred billion in the human brain, distinct functions and subtasks are presumably carried out by the interaction within and between relevant microcircuits. Within a neuronal network, there are two organizational topologies or connectivity classes:  structural and dynamic.  Structural connectivity describes the physical locations and connections between cells in the network, while dynamic connectivity is transient and changing and describes how one cell affects another. It is a subset of the physical connectivity \cite{Buibas:2011vu}. The neuroscience literature distinguishes two types of dynamic connectivity, functional connectivity, which describes statistical but not casual correlations between cells (or populations of cells and even entire brain regions), and effective connectivity, which is stronger and assumes casual dynamic connectivity within the network \cite{Bullmore:2009iv,Sporns:2004p1279,Sporns:2010p7996}.  In a neurobiological context, this means that cells that are physically connected need not necessarily signal each other. While there are major efforts aimed at mapping the structural and (electro)physiological connectivity of brain networks, such as the White House BRAIN Initiative aimed at the development of neurotechnologies capable of high density neural recordings, theoretical and computational methods for analyzing and making sense of all this data seem to be further behind. And while the experimental and technical challenges of imaging and mapping structural connectivity are tremendous, the objectives and goals are clear, and continued technical advances have ensured continued progress. Theoretically though it is much less obvious how to analyze, model, and use this data for the purposes of understanding brain function. One logical first step is to ask what is the effective connectivity associated with a specific imaged or recorded pattern of dynamic activity in a measured network or neural circuit with a given structural connectivity. Achieving this is at the forefront of theoretical and computational neuroscience, and certainly not a trivial task. Theoretically there are open mathematical problems that need to be solved in order to be able to accomplish this. Here, we propose and provide a summary of an approach for calculating effective connectivity from experimental observations of neuronal network activity. It builds on a neural signaling dynamic framework we have previously published \cite{:2011vu}. The proposed method operates on network-level data, makes use of all relevant prior knowledge, such as dynamical models of individual cells in the network and the physical structural connectivity of the network, and is broadly applicable to large classes of both biological and non-biological networks. 

Several relatively recently developed methods have attempted to tackle the problem of functional connectivity estimation, approaching the problem from different perspectives.  Eichler and colleagues developed a statistical-based approach that works with acyclic networks and computes direction and polarity of network connections \cite{Eichler:2003hd}.  Another method optimizes both neuronal and connectivity parameters of a simple linear integrate and fire deterministic model by comparing simulated output to experimental data, in this case being artificially generated spike trains \cite{Makarov:2005hf} .  The method was validated for small networks of five neurons or less, and can handle feedback connections, though the estimated connection strengths are only reliable for indicating polarity and not relative synaptic strength. deFeo's method uses three successive optimization steps to estimate a full state-space reconstruction of a measured signal, fitting of a local nonlinear dynamical model to the reconstructed signal, and then estimating a linear model to the interactions of the individual local models \cite{deFeo:2008km}. Finally, work by Eldawlatly, Zhou, Jin and Oweiss \cite{Eldawlatly:2010bo}, employs dynamic Bayesian networks for identification of connections of small ($N=10$) networks using a Poisson spiking model. More recently, Abarbanel and colleagues have described a powerful parameter estimation method optimized for tracking and predicting output voltages from unobservable internal state variables in models of dynamic neuronal signaling and coupled networks \cite{Abarbanel:2008,Abarbanel:2009ft}. However it remains unclear how their methods apply to and could take advantage of geometric networks where spatial information in addition to functional electrophysiological information is preserved. 

While these approaches are successful on the test data sets used for their validation, our proposed approach is broader and more applicable to experimentally derived observations.  Our approach is model agnostic, meaning that it can work with many different neuronal and astrocytic individual cell dynamics models.  The only constraint is that the dynamics are observable, though this constraint is valid for any type of estimation.  Our approach works with any type of network connectivity, meaning that a network can have recurrent connections (cyclic or acyclic).  Finally, this approach takes advantage of the linear summation of signals into a cell to compute both functional connections and signaling delays which are present in biological networks, providing a novel approach to estimating causal functional relationships in neuronal networks. We can also derive the minimum experimental data collection requirements, both in signal quantity and diversity, needed for mapping of a network of arbitrary size.

Our previous work on network mapping methods and algorithms depends on stochastic non-linear identification methods that take advantage of high performance massive parallel graphics processing unit (GPU) computing architectures \cite{:2011vu}. Although this approach is powerful, because of the dependency on stochastic numerical methods it can show significant variation in accuracy as a function of the non-linear optimizer used or the available computing power. By accuracy, we mean how close a cost function can be minimized in order to recover the real underlying effective connectivity of the network for a given transient signaling event. While on-going research in both non-linear optimization and GPU computing will ensure that our current methods will only improve, a deterministic algebraic identification approach to appropriately constrained problems would ensure that as long as the constraints of any such method are satisfied, the accuracy of the mapped network will always be within arbitrary error bounds.


\section{Neural Cell Signaling in Geometric Networks}

We very briefly summarize our geometric network neural signaling framework. Uniquely, this framework takes advantage of the physical structure and connectivity of a neuronal network in addition the individual (electrophysiological) internal neuronal dynamics to determine and analyze the network dynamics. We refer the reader to \cite{Buibas:2011vu} for complete details. We define a network as the graph $G=(V,E)$ to be a collection of vertices $V$ and connectivity information $E$ between vertices.  The network is composed of $N$ vertices, defined both geometrically (physical position) and dynamically (time variant state space):
\begin{equation}
V=\{ \xb_{j},\yb_{j}(t):\quad j=1,\ldots,N\}
\end{equation}
Here $\xb_{j}$ denotes the physical position in space of the vertex, while $\yb_{j}(t)$ is the state vector of the vertex at time $t$.  Connectivity information is composed of a connective strength $\omega_{ij}$ and time delays $\tau_{ij}$, defined for all unique and directed pairs of vertices in the network:
\begin{equation}
E=\{\omega_{ij},\tau_{ij}:\quad i,j=1,\ldots,N,\quad i\neq j\}
\end{equation}
The connective strengths are scaled such that $-1\leq \omega_{ij} \leq 1$ and delays are non-negative: $0\leq \tau_{ij}$. We describe and validate the full framework in \cite{:2011vu}.

Within a vertex, the time course of the state vectors is computed from the time map $\Hb_{j}$ between evenly spaced and sufficiently close time points, defined as follows:
\begin{equation}
\yb_{j}(t+\Delta t)=\mathbf{H}_{j}(\yb_{j}(t),\mathbf{s}_{j}(t),\ub_{j}(t))
\end{equation}
Here, $\ub_{j}(t)$ are the known external inputs to the system, and $\mathbf{s}_{j}(t)$ represents the internal network signaling within the network. Relatively little restriction is placed on the functional form of $\Hb_{j}$;  it can be linear or nonlinear, stochastic or deterministic. $\Hb_{j}$ must be observable, meaning that all the states of $\yb_{j}$ can be estimated from observations of a subset of $\yb_{j}$. Several approaches exist for state estimation or filtering: particle filtering \cite{Doucet:2003vh}, Kalman filters\cite{Kolas:2009kj}, and variational approaches \cite{Abarbanel:2009ft}.  The choice of estimation algorithm will depend on the individual cell models, and a full description of each is beyond the immediate scope of this paper.

Intra-vertex network signaling must be of the form:
\begin{equation}
\mathbf{s}_{j}(t)=\sum_{i=1,i\neq j}^{N}\omega_{ij}g_{i}\left(\yb_{i}(t-\tau_{ij})\right)
\label{eq:sjt}
\end{equation}
Under this assumption, the incoming signal to a vertex $j$ is the sum of weighted functions of the delayed states of other vertices in the network.  $g_{i}(\cdot)$ is the transmission function that operates on the state vector of another vertex.

Finally, in similar fashion to the transmission function, we define an observation entity $\mathbf{z}_{j}(t)$ as some function of the state vector $\yb_{j}(t)$:
\begin{equation}
\mathbf{z}_{j}(t)=F_{j}(\yb_{j}(t))
\end{equation}
$\mathbf{z}_{j}(t)$ describes what is experimentally observable about the vertex $j$ at time $t$.  The state transition and observation functions $\Hb_{j}$ and $F_{j}$, must be observable so that estimation of signaling quantities is possible. 

\section*{Estimation with known or no delays}
When the delays are known or non-existent, equation \ref{eq:sjt} can be written as:
\begin{equation}
\mathbf{s}_{j}(t)=\sum_{i=1,i\neq j}^{N}\omega_{ij}r_{i}(t)
\end{equation}
where $r_{j}(t)=g_{i}\left(\yb_{i}(t-\tau_{ij})\right)$.  Under the system observability assumption, we can estimate both $\mathbf{s}_{j}(t)$ and $r_{i}(t)$ from the observed activity of individual cells $\mathbf{z}_{j}(t)$ at each of the $T$ time points, the functional connective weights $\omega_{ij}$ can be obtained using some simple algebraic manipulation.  We can expand the above equation to matrix notation as follows:
\begin{equation}
\left[ \begin{array}{c} \mathbf{s}_{j}(0) \\\mathbf{s}_{j}(\Delta t) \\ \vdots \\ \mathbf{s}_{j}((T-1)\Delta t) \end{array}\right ] =
\left[
  \begin{array}{ c c c c}
     r_{1}(0) & r_{2}(0) & \cdots & r_{N}(0) \\
     r_{1}(\Delta t) & r_{2}(\Delta t) & \cdots & r_{N}(\Delta t) \\
     &\vdots  \\
     r_{1}((T-1)\Delta t) & r_{2}((T-1)\Delta t) & \cdots & r_{N}((T-1)\Delta t)
  \end{array} \right]
  \left[ \begin{array}{c} \omega_{1j} \\ \omega_{2j} \\ \vdots \\ \omega_{Nj} \end{array}\right ]
\end{equation}
The above equation can equivalently be written as:
\begin{equation}
\mathbf{S}_{j}=\mathbf{R}_{j}\mathbf{\omega}_{j}
\label{eq:linmat}
\end{equation}
The vector $\mathbf{S}_{j}$ has dimension $T \times 1$, the matrix $\mathbf{R}_{j}$ is $T\times (N-1)$ and the vector $\mathbf{\omega}_{j}$, containing all the incoming functional weights into vertex $j$ is thus $(N-1) \times 1$.  For a unique solution to $\mathbf{\omega}_{j}$, two criteria must be met: $T>N-1$ (ideally $T\gg N-1$), and $\mathbf{R}_{j}$ must have full rank (rank $N-1$), meaning that no two rows of  $\mathbf{R}_{j}$ can be linear multiples of each other.  Since equation \ref{eq:linmat} is linear, any standard method of factoring out $\omega_{j}$ can be used.  For example, one can multiply both sides by the pseudo-inverse of $\mathbf{R}_{j}$, isolating $\omega_{j}$ to the right hand side.

\section*{Estimation with unknown delays}
The more interesting and difficult case is when the delays of equation \ref{eq:sjt} are unknown.  To address this problem, we first take the Fourier transform of \ref{eq:sjt} as follows:
\begin{subequations}
\begin{equation}
\mathcal{F} \left\{ \mathbf{s}_{j}(t) \right\} =\mathcal{F} \left\{ \sum_{i=1,i\neq j}^{N}\omega_{ij}r_{i}(t) \right \}
\end{equation}
\begin{equation}
\mathbf{s}_{j}(s) = \sum_{i=1,i\neq j}^{N} \omega_{ij}r_{i}(s)e^{-2 \pi i\tau_{ij}s}
\label{eq:ffsjt}
\end{equation}
\end{subequations}
In the frequency domain, equation \ref{eq:ffsjt} employs the shift theorem, where a shift in the time domain of $\tau_{ij}$ results in a multiplication in the frequency domain by $e^{-2 \pi \tau_{ij}s}$.  This property allows us to effectively remove the unknown delay $\tau_{ij}$ from the argument of the $r_{i}$ function and express it as the argument of the natural exponent.  Since the exponent in \ref{eq:ffsjt} has only an imaginary argument, the equation can be expressed as follows:

\begin{equation}
\mathbf{s}_{j}(s) = \sum_{i=1,i\neq j}^{N} \omega_{ij}r_{i}(s)\left( \cos(-2 \pi \tau_{ij}s) + i \sin(-2 \pi \tau_{ij}s) \right)
\label{eq:ffsjtsc}
\end{equation}

We can now separate the real and imaginary components of the above, as
\begin{equation}
\text{Re}[\mathbf{s}_{j}(s)] +i\text{Im} [\mathbf{s}_{j}(s)]= \sum_{i=1,i\neq j}^{N} \omega_{ij}\left( \text{Re}[r_{i}(s)] + i\text{Im}[r_{i}(s)]\right)\left( \cos(-2 \pi \tau_{ij}s) + i \sin(-2 \pi \tau_{ij}s) \right)
\end{equation}
The above complex-valued equation can be separated into the real and imaginary parts, into two real-valued sets of equations:
\begin{subequations}
\begin{equation}
\text{Re}[\mathbf{s}_{j}(s)]= \sum_{i=1,i\neq j}^{N} \omega_{ij}\left( \text{Re}[r_{i}(s)]\cos(-2 \pi \tau_{ij}s)-\text{Im}[r_{i}(s)]\sin(-2 \pi \tau_{ij}s) \right)
\end{equation}
\begin{equation}
\text{Im} [\mathbf{s}_{j}(s)]= \sum_{i=1,i\neq j}^{N} \omega_{ij}\left( \text{Re}[r_{i}(s)]\sin(-2 \pi \tau_{ij}s)+\text{Im}[r_{i}(s)]\cos(-2 \pi \tau_{ij}s) \right)
\end{equation} \label{eq:risys}
\end{subequations}
We now have have a system of real-valued equations with $\omega_{ij}$ and $\tau_{ij}$ as unknowns and known values of $\mathbf{s}_{j}(s)$ and $r_{i}(s)$ evaluated at every frequency $s$.  The system is non-linear but can be solved using the Levenbergh-Marquardt solver (see \cite{Nocedal:2006ti} for description).  Briefly, one can rewrite equations \ref{eq:risys} as a series of functions that equal 0, and find that values of $\omega_{ij}$ and $\tau_{ij}$ that make the above equations zero or bring them closest to zero.

\begin{figure}
\begin{center}
\includegraphics[width=6in]{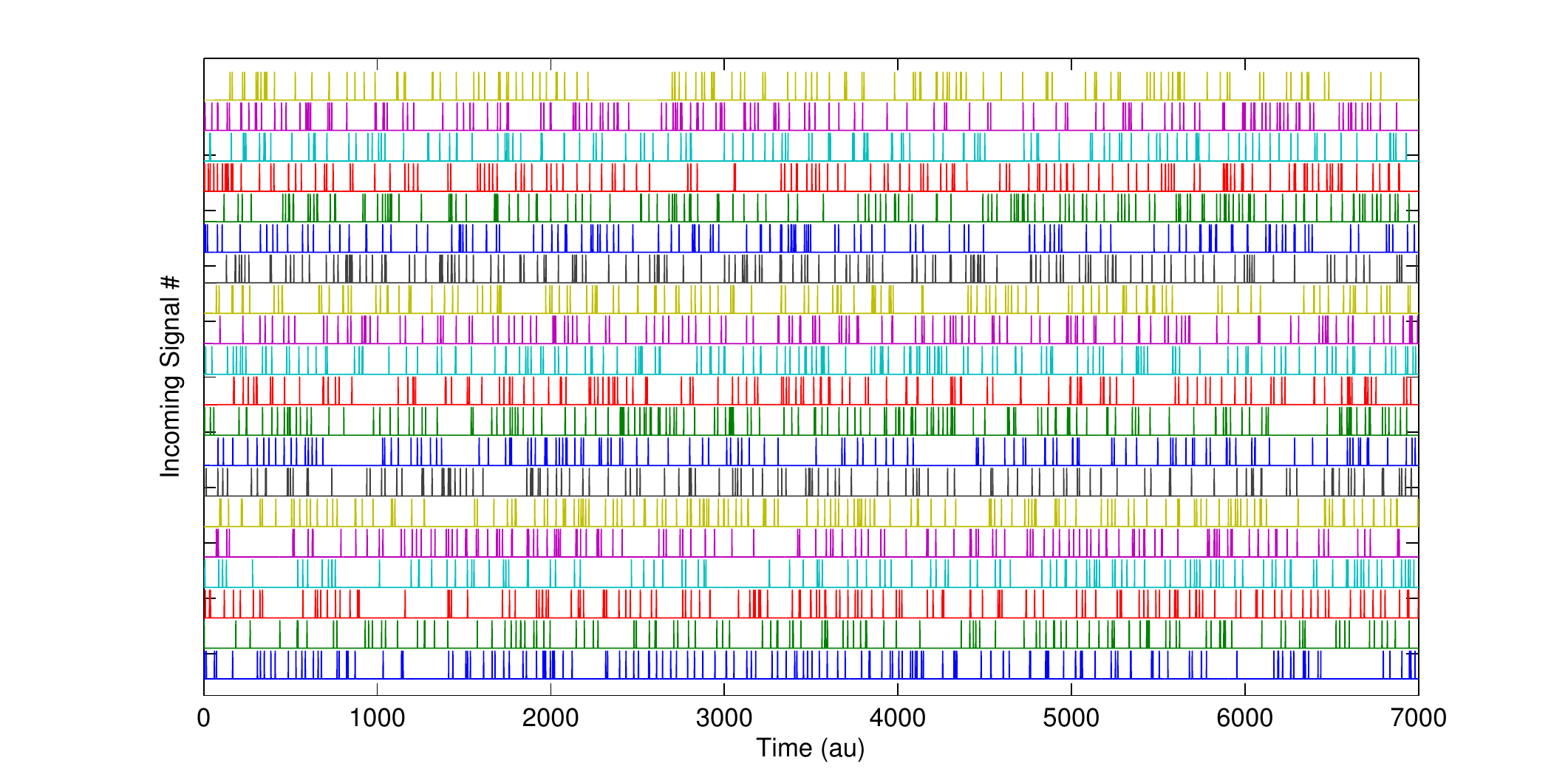}
\includegraphics[width=6in]{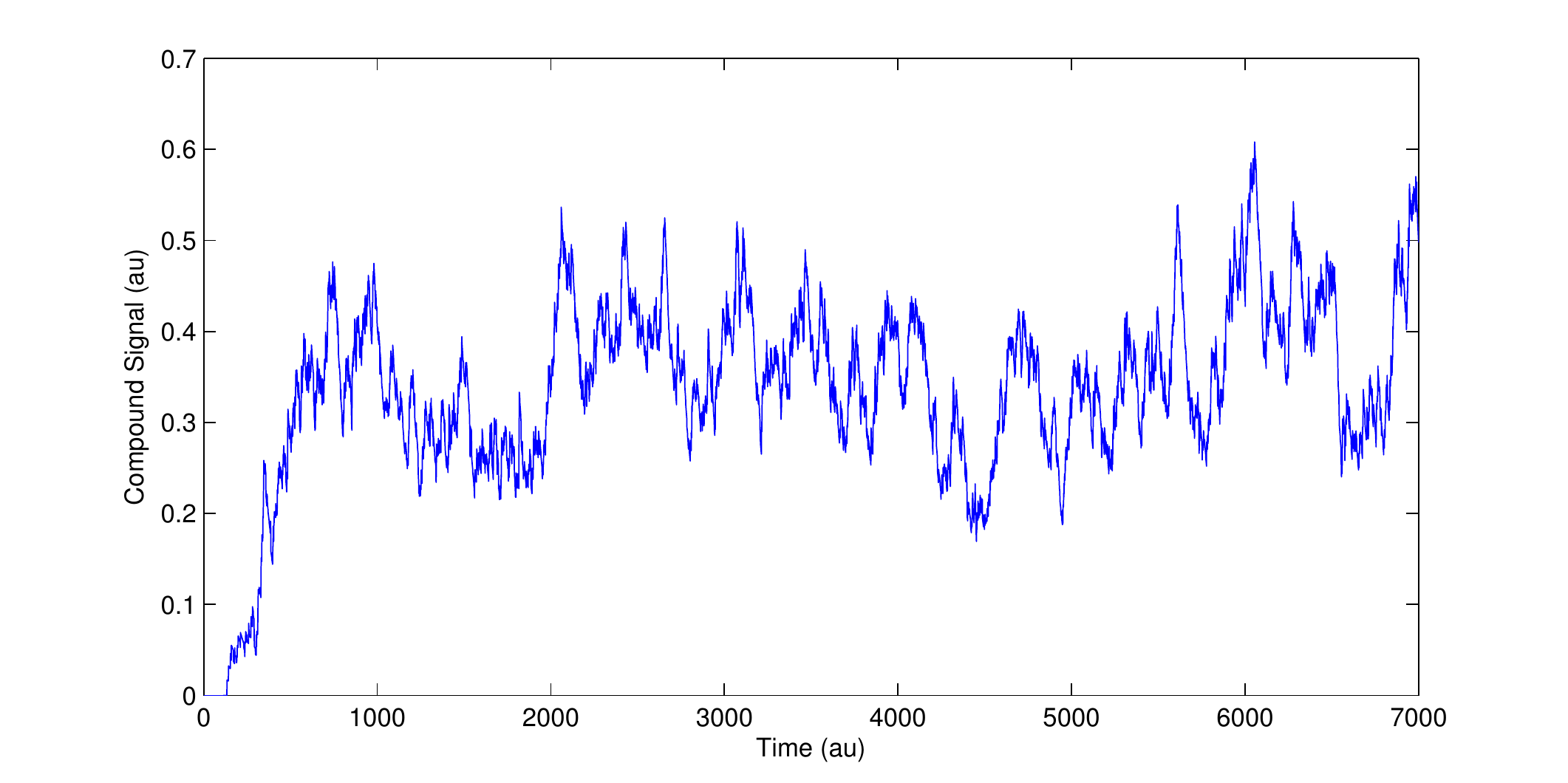}
\end{center}
\caption{\textbf{Incoming and Compound Measured Signals.} Measured transmitted (top panel) and compound signals into a vertex (bottom panel). The top figure shows the network transmitted signaling activity, while the bottom shows the driving compound signal into one vertex. Both of these are estimated from the observed activity of the individual vertices.
}
\label{fig:signals}
\end{figure}

\begin{figure}
\begin{center}
\includegraphics[width=6.5in]{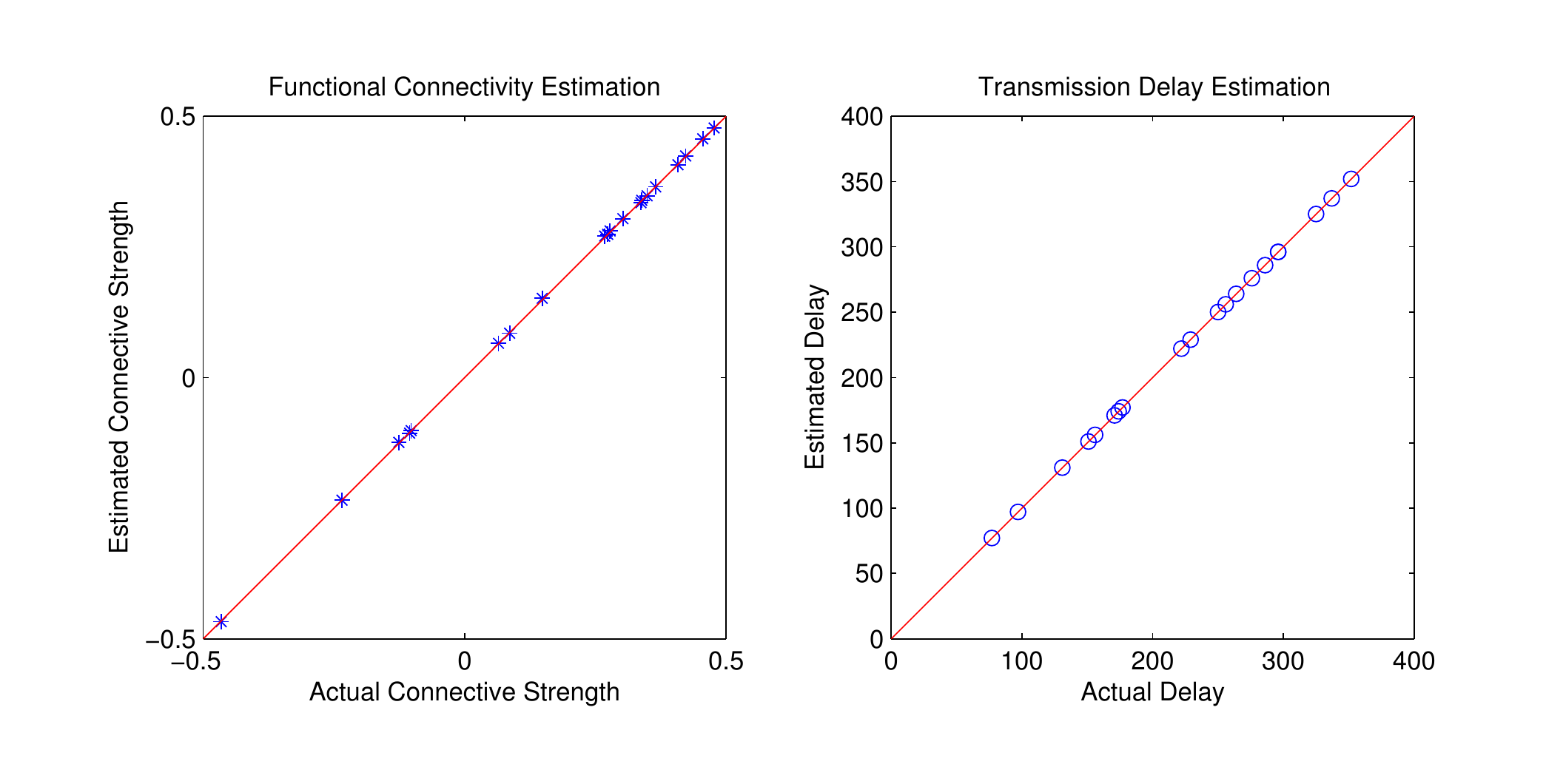}
\end{center}
\caption{\textbf{Connectivity and Delay Estimation.} Connectivity and Delays estimation for signals in Fig. 1. The system was solved using the Levenberg- Marquardt algorithm for nonlinear systems of equations. In this example, we estimated exactly the connective strengths and delays of the signals into the observed compound signal. 
}
\label{fig:Results}
\end{figure}

As an example, Fig. \ref{fig:signals} shows a simulated example of transmitted signals in a network, and the measured summed and delayed activity of one vertex. Once the transmitted signals are estimated, the individual contribution (i.e. effective connectivity) and delays from each transmitted signal $r_{i}(t)$ into the compound signal $\mathbf{s}_{j}(t)$ can be estimated by solving equation in \ref{eq:risys}.  The results are shown in Fig. \ref{fig:Results}.  In this example, we estimate 20 incoming functional connections and transmission delays into a vertex.  To generate the synthetic data, a random spike train was generated for each cell as the transmitted signal, as shown in figure \ref{fig:signals}A.  Dynamic connections were randomly selected in a uniform distribution between -0.5 and 0.5, and delays were randomly chosen between 0 and 400 time steps.  The incoming signals were weight-modulated and delayed to produce the incoming signal $\mathbf{s}_{j}(t)$ shown in the bottom panel of Fig. 1. Using the transmitted signals and the incoming signals, our algorithm was able to fully recover the original weights and delays that were used to generate the synthetic data (without the algorithm being aware or told what those were, of course).

\subsection*{Discussion}
This approach rests on the fundamental assumption that signals into a cell are summed and subsequently drive the dynamics of the cell in the network.  This assumption generally holds true for most neuronal and astrocytic networks, where post-synaptic currents and extra-cellular ATP are the signaling quantities that sum up and drive the cell dynamical systems.  From this assumption, we were able to express signaling in matrix form and perform the inversion with trivial matrix factorization (equation \ref{eq:linmat}).  What is apparent from this equation is that the number of time points collected from each cell has to be greater than the number of cells in the network for a unique solution to be computable.  Additionally, the outgoing signals of other cells in the network must be different so that no two cells have the same exact dynamics, so that no two rows of the matrix $\mathbf{R}_{j}$ are identical, making $\mathbf{R}_{j}$ singular and non-invertible. When delays are present and significant, the problem is tractable, though requires a non-linear least squares solver to compute.  An additional data requirement is that the individual outgoing signal differ in the frequency domain as well as in the time domain, meaning that two signals cannot be time-shifted copies of each other for a unique delay to be computable. Overall, these are necessary conditions, with sufficiency only possible for the linear, no-delay case.  In the non-linear, delays case, only necessary conditions can be met; the non-linearities in the summation of sines and cosines make the problem more difficult, mandating use of an optimizer.  Indeed few non-linear root-solving problems have sufficiency criteria for finding globally optimal solutions, as in the linear case.  While not unique to this problem, nonlinear root finding is the major theoretical obstacle in this algorithm.

Experimentally, the estimation approach we introduce here is tailored toward a mixed voltage and calcium recording experimental setup for neuronal functional connectivity.  Membrane voltage recordings from a few neurons at a time can be used to estimate the applied synaptic current, while calcium fluorescence observations of the complete network provides estimates of the spike times and thus the possible transmitted signals.  Combined, the incoming functional connections and delays are computed for the neurons whose voltage time courses are being recorded. 

\newpage
\bibliographystyle{amsplain}
\bibliography{Buibas&Silva.bib}

\end{document}